\RequirePackage{fixltx2e}
\documentclass[a4paper,fleqn,reqno,english,journal=cmatex,manuscript=article]{revtex4-2}
\usepackage[T1]{fontenc}
\usepackage[latin9]{inputenc}
\usepackage{color}
\usepackage{babel}
\usepackage{array}
\usepackage{float}
\usepackage{booktabs}
\usepackage{multirow}
\usepackage{amsmath}
\usepackage{amsthm}
\usepackage{amssymb}
\usepackage{graphicx}
\usepackage[unicode=true,pdfusetitle,
 bookmarks=true,bookmarksnumbered=false,bookmarksopen=false,
 breaklinks=false,pdfborder={0 0 1},backref=false,colorlinks=false]
 {hyperref}

\makeatletter

\providecommand{\tabularnewline}{\\}

\usepackage{lmodern}

\makeatother

\begin{document}
\title{Decoding optoelectronic behavior in X$_{3}$BI$_{3}$ antiperovskite
derivatives through many-body perturbation theory}
\author{Ayan Chakravorty}
\affiliation{Department of Physics, School of Natural Sciences, Shiv Nadar Institution
of Eminence, Greater Noida, Gautam Buddha Nagar, Uttar Pradesh 201314,
India}
\author{Surajit Adhikari}
\email{sa731@snu.edu.in}

\affiliation{Department of Physics, School of Natural Sciences, Shiv Nadar Institution
of Eminence, Greater Noida, Gautam Buddha Nagar, Uttar Pradesh 201314,
India}
\author{Priya Johari}
\email{priya.johari@snu.edu.in}

\affiliation{Department of Physics, School of Natural Sciences, Shiv Nadar Institution
of Eminence, Greater Noida, Gautam Buddha Nagar, Uttar Pradesh 201314,
India}
\begin{abstract}
Antiperovskite derivatives have emerged as promising candidates for
optoelectronic applications. However, due to the significant computational
cost, their excitonic and polaronic properties remain underexplored
despite being critical for optoelectronic performance. Here, we present
the structural, electronic, optical, excitonic, and polaronic properties
of a series of antiperovskite derivatives with the chemical formula
X$_{3}$BI$_{3}$ (X = Ca, Sr; B = P, As, Sb, Bi) using state-of-the-art
first-principles calculations. All the compounds exhibit direct bandgaps
with G$_{0}$W$_{0}$@PBE bandgap ranging from 2.42 to 3.02\,eV, optimal
for efficient light absorption with minimal energy loss. Exciton binding
energies (0.258$-$0.318\,eV) indicate moderate Coulomb attraction,
favoring exciton dissociation. Employing the Feynman polaron model,
we established the polaronic properties, where weak to intermediate
carrier-phonon coupling was observed, with polaron mobilities reaching
values up to 37.19 cm$^{2}$V$^{-1}$s$^{-1}$. These properties establish
X$_{3}$BI$_{3}$ materials as viable candidates for next-generation
optoelectronic devices.
\end{abstract}
\maketitle

\section{Introduction:}

Halide perovskites have garnered tremendous interest in optoelectronics,
due to their excellent light absorption, tunable bandgaps, and high
power conversion efficiencies \citep{C1-31,C1-32}. Yet, concerns
over lead toxicity and instability from volatile organic cations hinder
their practical use \citep{C1-33,C1-34}. To address this, several
Pb-free alternatives-such as chalcogenide perovskites, double perovskites
and vacancy-ordered perovskites have been explored, though many face
drawbacks like indirect bandgaps and complex synthethis. More recently,
halide antiperovskite derivatives with the formula X$_{3}$BA$_{3}$
(X = Ca, Sr; B = P, As, Sb, Bi; A= F, Cl, Br, I) have emerged as promising,
environmentally benign, and structurally robust candidates \citep{C2-8,C2-9}.
Their potential has been validated by successful syntheses, highlighting
their relevance for next-generation optoelectronic applications \citep{C2-1,C2-29}.

Structurally, halide antiperovskite derivatives (X$_{3}$BI$_{3}$)
can be viewed as an evolution from traditional perovskites (ABX$_{3}$)
and antiperovskites (X$_{3}$BA). In perovskites (ABX$_{3}$), A and
B are cations while X is an anion while in antiperovskites (X$_{3}$BA),
the ionic roles are inverted-X becomes a cation, while B and A are
anionic species \citep{C2-28}. Extending this framework, halide antiperovskite
derivatives introduce three halide anions (3 $\times$ A$^{-}$) in
place of the single anion \textquotedblleft A$^{3-}$\textquotedblright{}
to satisfy charge neutrality. This structural modification not only
preserves the cubic symmetry but also provides added flexibility to
tune electronic and optical properties for efficient optoelectronic
functionality.

Numerous studies have recently explored the potential of halide antiperovskite
derivatives for optoelectronic applications, highlighting their promising
electronic and optical characteristics \citep{C2-10,C2-11,C2-12,C2-13,C2-14,C2-15}.
For example, Sr$_{3}$PnCl$_{3}$ (Pn = P, As, Sb) and A$_{3}$NCl$_{3}$(A
= Ba, Sr, Ca) exhibit direct bandgaps in the range of 1.95$-$2.14\,eV
and 0.58$-$1.68 eV, alongside strong absorption and mechanical robustness,
highlighting their suitability for solar energy harvesting \citep{C2-8,C2-9}.
Joifullah et al. investigated the pressure-dependent structural, electronic,
mechanical, and optical behavior of Sr$_{3}$PX$_{3}$ (X = Cl, Br),
confirming their semiconducting nature and enhanced optoelectronic
properties under strain \citep{C2-16}; however, experimental validation
for these systems remains absent.

Experimental reports on iodide-based antiperovskite derivatives have
further inspired the focus on these materials \citep{C2-1}. Several
studies have explored their structural, electronic, and optical behavior
\citep{C2-19,C2-21,C2-22,C2-23,C2-24,C2-25,C2-26}.\textcolor{red}{{}
}\textcolor{black}{Notably, Rahman et al. investigated Ba$_{3}$PI$_{3}$,
Ba$_{3}$AsI$_{3}$, and Ba$_{3}$SbI$_{3}$, reporting direct bandgaps
($\sim$ 1.9$-$2.3\,eV), strong absorption, and simulated PCEs exceeding
21$-$29 \% via SCAPS-1D, while Liu et al. assessed Ba$_{3}$MX$_{3}$
(M = As, Sb; X = Cl, Br, I), revealing direct bandgaps (1.35$-$1.65\,eV),
low effective masses, and theoretical efficiencies up to 31.9 \%}
\citep{C2-17,C2-18}. Although these studies lay important groundwork,
key optoelectronic properties-such as excitonic and polaronic response,
remain unaddressed, highlighting the need for deeper theoretical insight.

To address this gap, here we present a comprehensive study of the
structural, electronic, optical, excitonic, and polaronic properties
of halide antiperovskite derivatives X$_{3}$BI$_{3}$ (X = Ca, Sr;
B = P, As, Sb, Bi), all crystallizing in the cubic Pm$\bar{3}$m phase.
By employing density functional theory (DFT) \citep{C1-57,C1-58},
hybrid functional (HSE06) \citep{C1-19}, density functional perturbation
theory (DFPT) \citep{C1-59}, and many-body perturbation theory (GW-BSE)
based simulations \citep{C1-60,C1-61}, we explore the overall optoelectronic
potential of these materials. Our calculations reveal that 5 out of
8 compositions are dynamically stable at 0\,K, with G$_{0}$W$_{0}$@PBE
bandgaps ranging from 2.46 to 3.02\,eV, falling well within the visible
spectrum. Exciton binding energies obtained via the BSE formalism
lie between 0.258$-$0.318\,eV, and polaron mobility reaches values
as high as 26.23\,cm$^{2}$V$^{-1}$s$^{-1}$ for electrons and 37.19\,cm$^{2}$V$^{-1}$s$^{-1}$
for holes, indicating promising charge transport characteristics.
Computational details are provided in the Supplemental Material.

\section{Results and Discussions:}

We conduct an in-depth and systematic exploration of halide antiperovskite
derivatives X$_{3}$BI$_{3}$ (X = Ca, Sr; B = P, As, Sb, Bi) to assess
their suitability for optoelectronic applications. The subsequent
sections provide a comprehensive evaluation of their structural and
dynamical stability, followed by detailed insights into their electronic,
optical, excitonic, and polaronic behavior, offering a solid theoretical
foundation for guiding future experimental endeavors.

\subsection{\textit{Crystal Structure and stability:}}

\begin{figure}[H]
\begin{centering}
\includegraphics[bb=0bp 0bp 952.582bp 464.899bp,width=1\textwidth,height=1\textwidth,keepaspectratio]{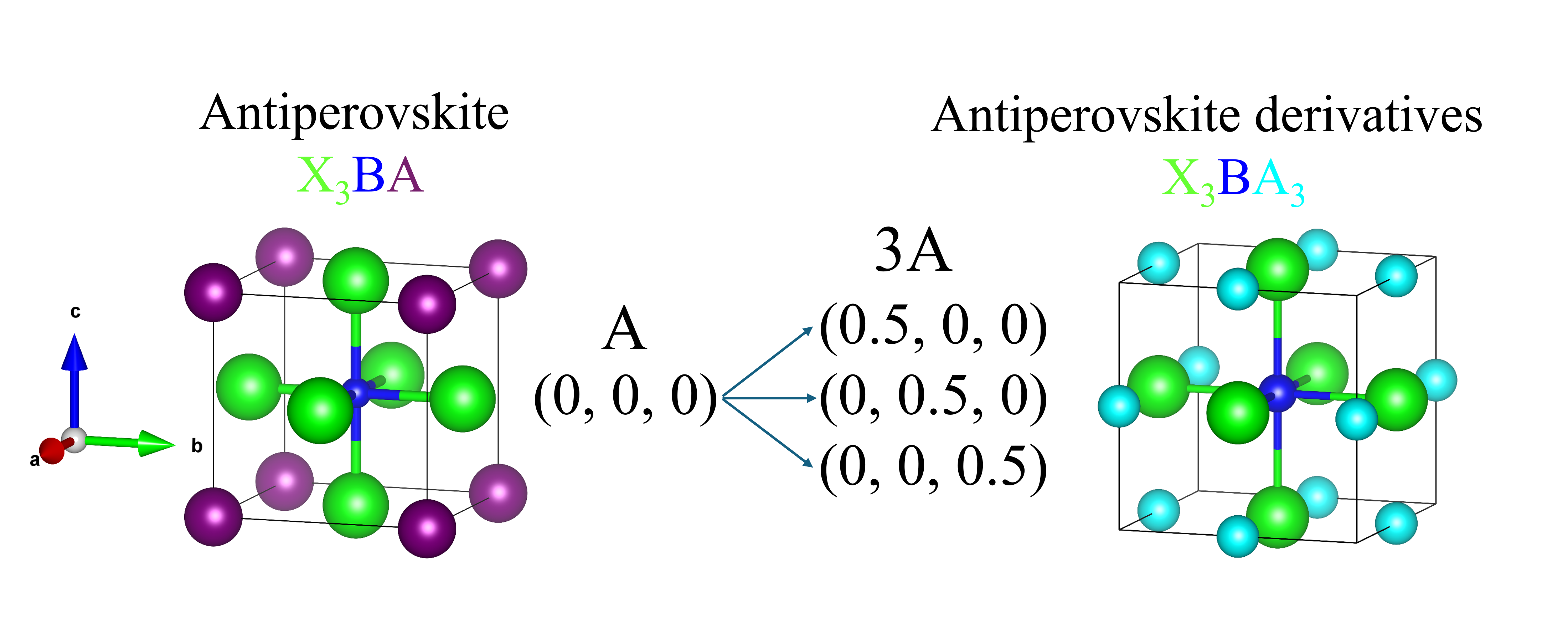}
\par\end{centering}
\caption{\label{fig:1}Schematic illustration of the transformation from antiperovskite
X$_{3}$BA (left) to its derivative X$_{3}$BA$_{3}$ (right) structure.
In X$_{3}$BA, the single A atom at (0, 0, 0) is replaced in X$_{3}$BA$_{3}$
by three A atoms located at (0.5, 0, 0), (0, 0.5, 0), and (0, 0, 0.5).
The color scheme in the chemical formulas corresponds to the atom
colors in the structures: green- X-site atoms (Ca/Sr), blue- B-site
atoms (P/As/Sb/Bi), purple- \textquotedblleft A\textquotedblright{}
atom in X$_{3}$BA, and cyan- three \textquotedblleft A\textquotedblright{}
atoms in X$_{3}$BA$_{3}$. In the present study, we specifically
investigate the X$_{3}$BA$_{3}$ systems with X = Ca, Sr; B = P,
As, Sb, Bi; and A = I.}
\end{figure}

To examine the structural and dynamical stability of the X$_{3}$BI$_{3}$
(X = Ca, Sr; B = P, As, Sb, Bi) compounds, we analyze their optimized
crystal structures, as illustrated in Fig. \ref{fig:1}, and assess
key geometric trends in Table \ref{tab:1}. All the materials crystallize
in the cubic crystal structure (space group - Pm$\bar{3}$m, No. 221)
as revealed in experimental observations \citep{C2-1}. The unit cell
contains seven atoms, with Ca/Sr at 3c (0.5, 0, 0.5), P/As/Sb/Bi at
1b (0.5, 0.5, 0.5), and I at 3d (0, 0.5, 0). A general expansion of
lattice parameters and bond lengths is observed toward the heavier
pnictogens (P/As/Sb/Bi), but the trend is not strictly monotonic.
In Ca$_{3}$BI$_{3}$ series, the lattice constant is 6.25 $\textrm{\AA}$
for both Ca$_{3}$PI$_{3}$ and Ca$_{3}$AsI$_{3}$, rising to 6.41
$\textrm{\AA}$ for Ca$_{3}$SbI$_{3}$; the X-I and X-B bond lengths
increases from 3.12 $\textrm{\AA}$ (Ca$_{3}$PI$_{3}$) to 3.21 $\textrm{\AA}$
(Ca$_{3}$SbI$_{3}$) with only marginal change for Ca$_{3}$BiI$_{3}$.
The Sr$_{3}$BI$_{3}$ series follows the same overall increase (a
= 6.52 $\rightarrow$ 6.56 $\rightarrow$ 6.73 $\textrm{\AA}$) with
Bi nearly identical to Sb, and bond lengths (3.26 $\rightarrow$ 3.28
$\rightarrow$ 3.37 $\textrm{\AA}$) showing similar small increments.
At fixed B, Sr-based compounds consistently have larger lattice constants
and bond lengths than Ca analogues, reflecting the larger ionic radius
of Sr$^{2+}$.

\begin{figure}[H]
\begin{centering}
\includegraphics[bb=0bp 0bp 952.637bp 464.791bp,width=1\textwidth,height=1\textwidth,keepaspectratio]{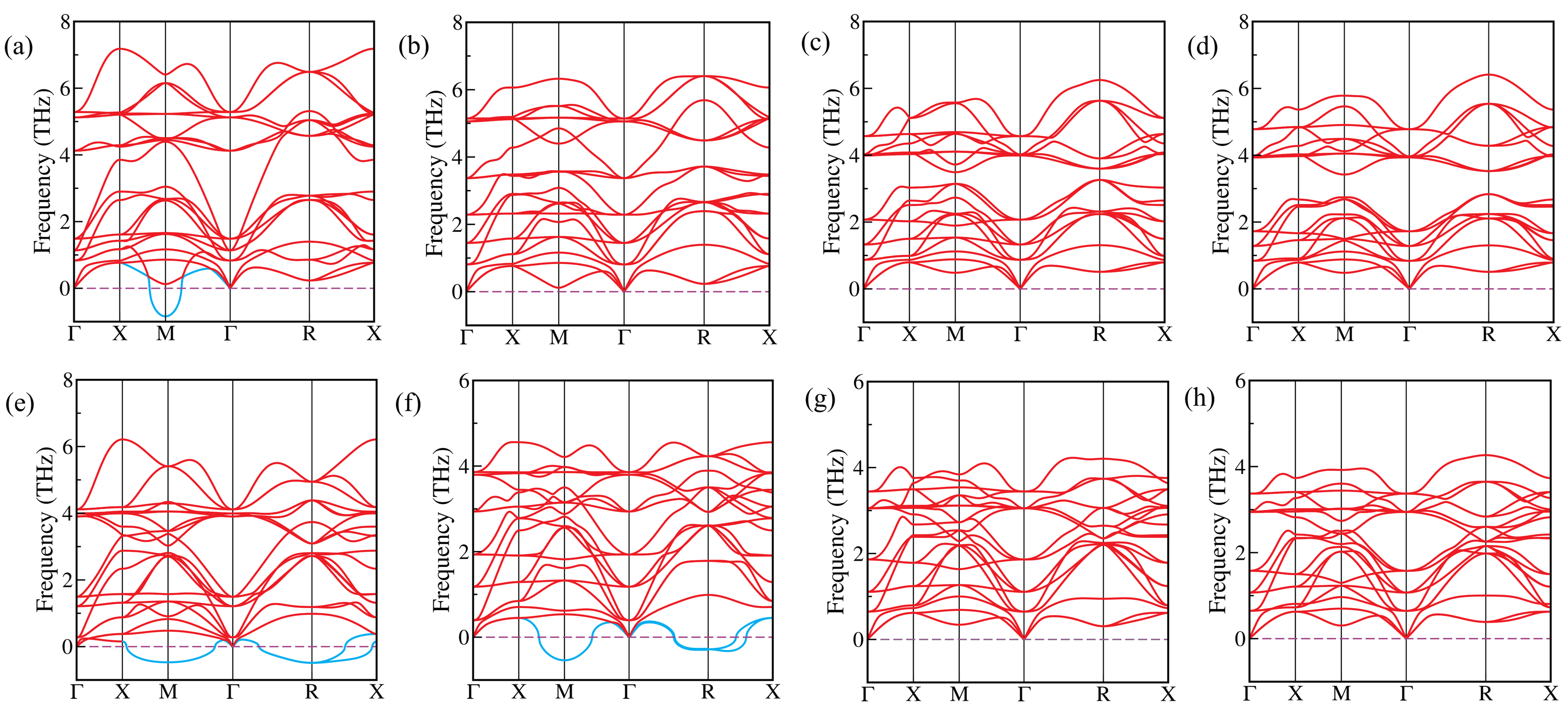}
\par\end{centering}
\caption{\label{fig:2}Phonon dispersion curves of the a) Ca$_{3}$PI$_{3}$,
b) Ca$_{3}$AsI$_{3}$, c) Ca$_{3}$SbI$_{3}$, d) Ca$_{3}$BiI$_{3}$,
e) Sr$_{3}$PI$_{3}$, f) Sr$_{3}$AsI$_{3}$, g) Sr$_{3}$SbI$_{3}$,
and h) Sr$_{3}$BiI$_{3}$ antiperovskite derivative, respectively.}
\end{figure}

To further validate the stability of these phases, we evaluate their
phonon dispersion curves using density functional perturbation theory
(DFPT) \citep{C1-59}, as shown in Fig. \ref{fig:2}. The absence
of imaginary frequencies across the Brillouin zone confirms that Ca$_{3}$AsI$_{3}$,
Ca$_{3}$SbI$_{3}$, Ca$_{3}$BiI$_{3}$, Sr$_{3}$SbI$_{3}$, and
Sr$_{3}$BiI$_{3}$ are dynamically stable in the cubic phase at 0
K. In contrast, Ca$_{3}$PI$_{3}$, Sr$_{3}$PI$_{3}$, and Sr$_{3}$AsI$_{3}$
display negative frequencies, signaling dynamic instabilities at 0
K. These instabilities are likely associated with spontaneous symmetry
breaking toward lower-symmetry distorted phases at low temperatures.
However, such structures may still be stabilized at elevated temperatures,
as commonly observed in many experimental cases.

To assess thermodynamic stability, the decomposition enthalpy for
each X$_{3}$BI$_{3}$ compound is calculated based on the most likely
binary decomposition pathways \citep{C2-2} (For details, see the
Supplemental Material). All materials exhibit positive decomposition
enthalpy values (see $\mathrm{\Delta H}$ values in Table \ref{tab:1}),
indicating energetic favorability of the X$_{3}$BI$_{3}$ phase over
its binary constituents. In addition to the dynamic and thermodynamic
stability, the mechanical stability and elastic properties of the
systems are also studied by calculating second ordered elastic constants
and all the materials are found to be mechanically stable (detalied
in the Supplemental Material).
\begin{center}
\begin{table}[H]
\caption{\label{tab:1}Calculated lattice parameters, bond lengths, and decomposition
energies of X$_{3}$BI$_{3}$ (X = Ca, Sr; B = P, As, Sb, Bi) antiperovskite
derivatives.}

\centering{}%
\begin{tabular}{ccccccccccc}
\hline 
\multirow{2}{*}{Configurations} & \multirow{2}{*}{} & \multirow{2}{*}{} & \multicolumn{3}{c}{Lattice parameter (a = b = c)} & \multirow{2}{*}{} & \multirow{2}{*}{} & Bond length & \multirow{2}{*}{} & Decomposition energy\tabularnewline
\cline{4-6} \cline{5-6} \cline{6-6} \cline{9-9} \cline{11-11} 
 &  &  & Our study ($\textrm{\AA}$) &  & Previous reports ($\textrm{\AA}$) &  &  & X$-$I and X$-$B ($\textrm{\AA}$) &  & $\mathrm{\Delta H}$ (eV/atom)\tabularnewline
\hline 
Ca$_{3}$PI$_{3}$ &  &  & 6.25 &  & 6.20 \citep{C2-24} &  &  & 3.12 &  & 0.085\tabularnewline
Sr$_{3}$PI$_{3}$ &  &  & 6.52 &  &  &  &  & 3.26 &  & 0.140\tabularnewline
Ca$_{3}$AsI$_{3}$ &  &  & 6.25 &  & 6.27 \citep{C2-19}, 6.25 \citep{C2-26} &  &  & 3.13 &  & 0.111\tabularnewline
Sr$_{3}$AsI$_{3}$ &  &  & 6.56 &  &  &  &  & 3.28 &  & 0.157\tabularnewline
Ca$_{3}$SbI$_{3}$ &  &  & 6.41 &  & 6.41 \citep{C2-30} &  &  & 3.21 &  & 0.150\tabularnewline
Sr$_{3}$SbI$_{3}$ &  &  & 6.73 &  &  &  &  & 3.37 &  & 0.185\tabularnewline
Ca$_{3}$BiI$_{3}$ &  &  & 6.41 &  & 6.38 \citep{C2-25} &  &  & 3.20 &  & 0.135\tabularnewline
Sr$_{3}$BiI$_{3}$ &  &  & 6.72 &  & 6.69 \citep{C2-25} &  &  & 3.36 &  & 0.166\tabularnewline
\hline 
\end{tabular}
\end{table}
\par\end{center}

\subsection{\textit{Electronic Properties:}}

After confirming the structural stability, electronic structure calculations
for X$_{3}$BI$_{3}$ compounds are performed, as these properties
are crucial to design a photoelectric device. In this context, the
partial density of states (PDOS), total density of states (TDOS),
band-edge positions, and the nature of the band gap are analyzed to
obtain comprehensive insights into the electronic structure.

Figure S1 illustrates the orbital-resolved density of states calculated
using HSE06 xc functional for the X$_{3}$BI$_{3}$ compounds. In
these systems, the valence band maximum (VBM) is primarily composed
of B-$p$ and I-$p$ orbitals, while the conduction band minimum (CBM)
is mainly composed of X-site $d$ orbitals (Ca-$3d$ or Sr-$4d$),
with minor I-$p$ contributions, supporting $p$-$d$ optical transitions.
Interestingly, although Ca-containing compounds show sharper peaks
near CBM, they possess wider band gaps compared to their Sr analogs.
This trend can be rationalized by considering the larger lattice parameters
and longer X$-$B and X$-$I bond lengths in Sr-based systems, which
induce weaker orbital overlap and increased delocalization, effectively
narrowing the band gap. Thus, despite sharper DOS peaks for Ca variants,
the broader Sr-based lattices facilitate band edge convergence. Unlike
conventional halide perovskites, here the alkaline-earth X-site cation
plays a non-trivial role in shaping the CBM due to the active participation
of its $d$ orbitals.

Band structure calculations were initially performed using the PBE
xc functional \citep{C1-17}, which is known to underestimate bandgaps
due to self interaction error. Therefore, to obtain more accurate
bandgap estimation, we employed the HSE06 hybrid functional and the
many-body perturbation theory (MBPT) based G$_{0}$W$_{0}$@PBE approach
\citep{C1-19,C1-20,C1-21}. Additionally, for the PBE calculations,
spin-orbit coupling (SOC) is explicitly considered, especially due
to the presence of heavy elements such as Bi, where relativistic effects
significantly influence band dispersion. Notably, the PBE-SOC bandgap
of Ca$_{3}$AsI$_{3}$ closely aligns with previous report \citep{C2-3},
supporting the reliability of our calculations.
\begin{center}
\begin{figure}[H]
\begin{centering}
\includegraphics[width=1\textwidth,height=1\textheight,keepaspectratio]{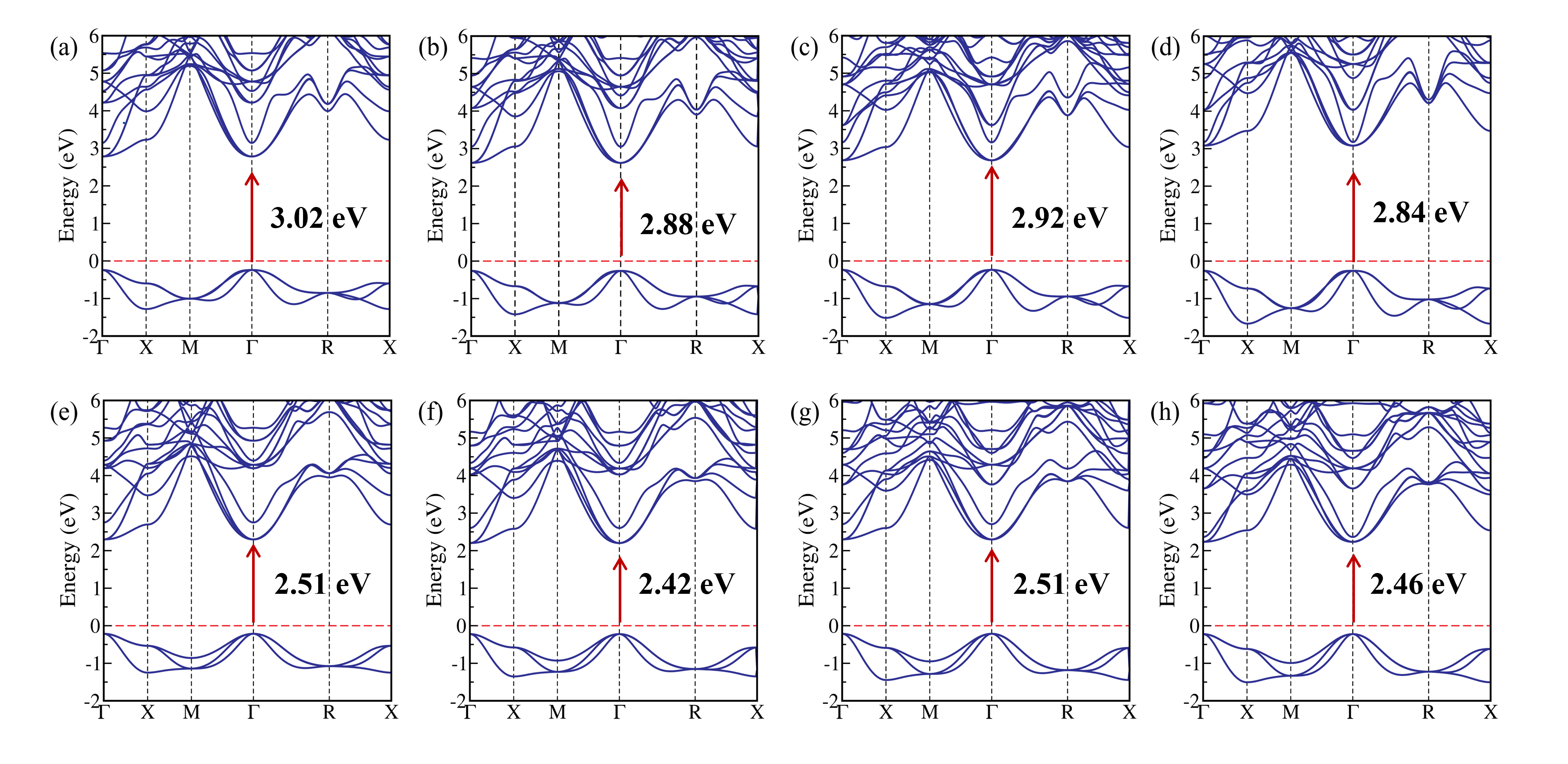}
\par\end{centering}
\caption{\label{fig:3}Electronic band structures of a) Ca$_{3}$PI$_{3}$,
b) Ca$_{3}$AsI$_{3}$, c) Ca$_{3}$SbI$_{3}$, d) Ca$_{3}$BiI$_{3}$,
e) Sr$_{3}$PI$_{3}$, f) Sr$_{3}$AsI$_{3}$, g) Sr$_{3}$SbI$_{3}$,
and h) Sr$_{3}$BiI$_{3}$ antiperovskite derivative, respectively,
calculated using the G$_{0}$W$_{0}$@PBE method. The fermi level
is set to be zero and marked by the dashed line.}
\end{figure}
\par\end{center}

All the studied X$_{3}$BI$_{3}$ systems (X = Ca, Sr; B = P, As,
Sb, Bi) exhibit direct bandgaps located at the $\Gamma$-point. The
valence band maximum (VBM) and conduction band minimum (CBM) consistently
occur at this high-symmetry k-point, suggesting favorable optical
transitions for light absorption and emission. The bandgap values
extracted from the G$_{0}$W$_{0}$@PBE (Figure \ref{fig:3}) method
show a consistent trend across the series (Table \ref{tab:2}), where
the bandgap decreases as we move from Ca- to Sr-based compounds for
a fixed B-site atom. For instance, Ca$_{3}$PI$_{3}$ shows a G$_{0}$W$_{0}$
bandgap of 3.02 eV, while Sr$_{3}$PI$_{3}$ drops to 2.51 eV, a trend
observable for the other B-site variants as well. This narrowing of
bandgap in Sr-based systems correlates with their larger lattice parameters
and longer bond lengths, which can induce enhanced orbital delocalization,
thereby reducing the bandgap as discussed earlier. Importantly, the
direct nature of the bandgap and its alignment with the visible spectrum
make these materials highly promising with minimal energy loss for
optoelectronic applications such as photovoltaics and LEDs.

\begin{table}[H]
\caption{\label{tab:2}Computed bandgaps ($E_{g}$) of X$_{3}$BI$_{3}$ (X
= Ca, Sr; B = P, As, Sb, Bi) antiperovskite derivatives using the
PBE/PBE-SOC, HSE06, and\textbf{ }G$_{0}$W$_{0}$@PBE method. Here,
$m_{e}^{*}$ is the electron effective mass, $m_{h}^{*}$ is the hole
effective mass, and $\mu^{*}$ is the reduced mass of charge carriers.
All values of the effective mass are in terms of free-electron mass
($m_{0}$).}

\centering{}%
\begin{tabular}{ccccccccc}
\hline 
\multirow{2}{*}{Configurations} & \multicolumn{3}{c}{Bandgap (eV) } & \multirow{2}{*}{} & \multirow{2}{*}{} & \multirow{2}{*}{$m_{e}^{*}$ ($m_{0}$)} & \multirow{2}{*}{$m_{h}^{*}$ ($m_{0}$)} & \multirow{2}{*}{$\mu^{*}$ ($m_{0}$)}\tabularnewline
\cline{2-4} \cline{3-4} \cline{4-4} 
 & PBE/PBE-SOC & HSE06 & G$_{0}$W$_{0}$@PBE &  &  &  &  & \tabularnewline
\hline 
Ca$_{3}$PI$_{3}$ & 1.37/1.33 & 2.25 & 3.02 &  &  & 0.548 & 0.644 & 0.296\tabularnewline
Sr$_{3}$PI$_{3}$ & 1.27 /1.24 & 2.02 & 2.51 &  &  & 0.514 & 0.627 & 0.282\tabularnewline
Ca$_{3}$AsI$_{3}$ & 1.33/1.23 & 1.40 & 2.88 &  &  & 0.541 & 0.572 & 0.278\tabularnewline
Sr$_{3}$AsI$_{3}$ & 1.26/1.18  & 2.00 & 2.42 &  &  & 0.514 & 0.569 & 0.270\tabularnewline
Ca$_{3}$SbI$_{3}$ & 1.37/1.19 & 2.23 & 2.92 &  &  & 0.561 & 0.497 & 0.264\tabularnewline
Sr$_{3}$SbI$_{3}$ & 1.34/1.17 & 2.08 & 2.51 &  &  & 0.533 & 0.513 & 0.261\tabularnewline
Ca$_{3}$BiI$_{3}$ & 1.28/0.82 & 2.09 & \textcolor{black}{2.84} &  &  & 0.557 & 0.444 & 0.247\tabularnewline
Sr$_{3}$BiI$_{3}$ & 1.27/0.83 & 1.96 & 2.46 &  &  & 0.526 & 0.462 & 0.246\tabularnewline
\hline 
\end{tabular}
\end{table}

The effective masses (see Table \ref{tab:2}) are generally close
to or below the free electron mass, suggesting promising transport.
Notably, in Ca$_{3}$SbI$_{3}$, Ca$_{3}$BiI$_{3}$, Sr$_{3}$SbI$_{3}$,
and Sr$_{3}$BiI$_{3}$, the hole masses ($m_{h}^{*}$) are smaller
than electron masses ($m_{e}^{*}$), favoring hole transport. Conversely,
P- and As-based compounds, having lower electron effective masses
than hole masses, are expected to exhibit enhanced electron mobility
and n-type transport behavior.

\subsection{\textit{Dielectric and excitonic properties:}}

\begin{figure}[H]
\begin{centering}
\includegraphics[width=1\textwidth,height=1\textheight,keepaspectratio]{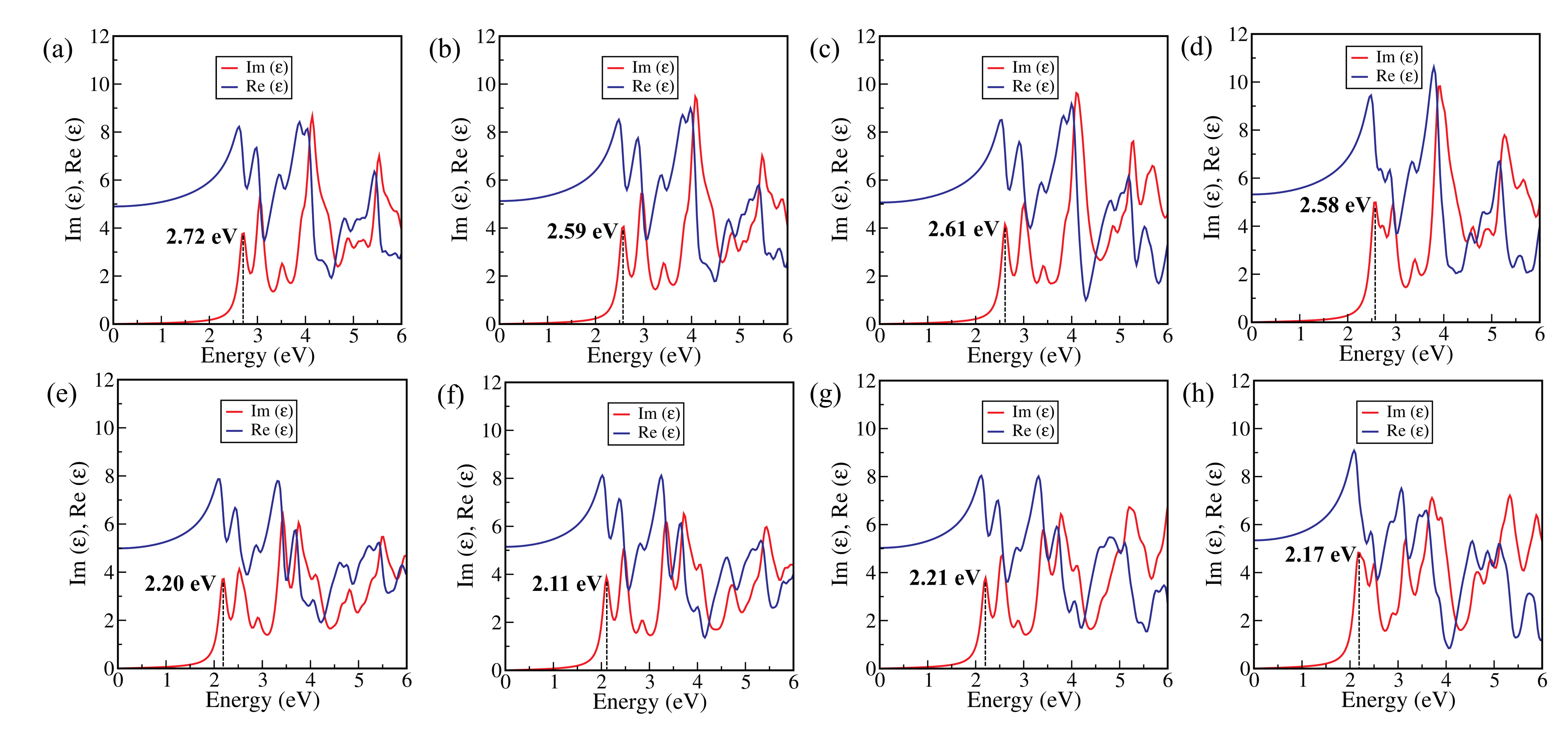}
\par\end{centering}
\caption{\label{fig:4}Real {[}Re ($\varepsilon$){]} and imaginary {[}Im ($\varepsilon$){]}
part of the dielectric function of a) Ca$_{3}$PI$_{3}$, b) Ca$_{3}$AsI$_{3}$,
c) Ca$_{3}$SbI$_{3}$, d) Ca$_{3}$BiI$_{3}$, e) Sr$_{3}$PI$_{3}$,
f) Sr$_{3}$AsI$_{3}$, g) Sr$_{3}$SbI$_{3}$, and h) Sr$_{3}$BiI$_{3}$
antiperovskite derivative, respectively, obtained using the BSE@G$_{0}$W$_{0}$@PBE
method.}
\end{figure}

While the studied X$_{3}$BI$_{3}$ compounds demonstrate suitable
bandgaps and promising electronic properties, these alone do not ensure
optimal optoelectronic performance. A comprehensive understanding
of their optical response and exciton formation is necessary, as these
bound electron-hole states critically affect light absorption and
charge transport. Fig. \ref{fig:4} shows the real and imaginary parts
of the dielectric function, where the real part reflects material
polarization and the imaginary part denotes optical absorption. All
of the X$_{3}$BI$_{3}$ systems exhibit strong absorption beginning
in the visible region ($\sim$ 2.1$-$2.7 eV) and extending up to
$\sim$ 5 eV, covering both the visible and near-UV spectral ranges.
The absorption edges display an overall red-shift trend from P to
Bi, consistent with the reduction in bandgap, although small deviations
occur for intermediate pnictogens.

Exciton binding energy ($E_{B}$) refers to the energy required to
separate an electron-hole which plays a crucial role in optoelectronic
performance. In our study, $E_{B}$, is calculated using the difference
between the direct quasiparticle (G$_{0}$W$_{0}$) bandgap and the
first optical absorption peak from BSE calculation \citep{C1-18,C1-25}.
For all X$_{3}$BI$_{3}$ systems, $E_{B}$ ranges from 0.258 and
0.318\,eV (Table \ref{tab:3}), signifying moderately bound excitons
that support efficient exciton dissociation and charge transport.
A-site variation (Sr vs. Ca) leads to slightly lower $E_{B}$ in Ca-based
systems compared to their Sr counterparts confirming their superiority
in exciton dissociation. 

To support and contextualize the BSE results, we additionally apply
the simplified Wannier-Mott (WM) model \citep{C2-33,C2-35} to provide
insight into dielectric screening and exciton delocalization (for
details, see Supplemental Material). Although the WM model reproduces
general trends, its oversimplified assumptions of parabolic band dispersion
and isotropic screening limit its quantitative reliability, especially
for systems with complex band structures. Here, to evaluate $E_{B}$,
$\mathrm{\varepsilon_{eff}}$ must be determined first. When $E_{B}\gg\hbar\omega_{LO}$,
as observed in our systems, the electronic contribution to the dielectric
function ($\varepsilon_{\infty}$) dominates the dielectric screening,
making $\varepsilon_{\infty}$ a valid approximation for $\mathrm{\varepsilon_{eff}}$.
In such cases, the influence of lattice relaxation is negligible,
and the ionic contribution can be safely ignored \citep{C1-13,C1-18,C1-26,C2-4}.
We employed the DFPT method \citep{C1-59} to estimate the ionic contribution
to the dielectric function ($\varepsilon_{ion}$), and the electronic
contribution ($\varepsilon_{\infty}$) is estimated using the BSE@G$_{0}$W$_{0}$@PBE
method. The upper ($E_{Bu}$) and lower ($E_{Bl}$) bound to the exciton
binding energy is computed from the contribution of electronic dielectric
constant ($\varepsilon_{\infty}$) and static dielectric constant
($\varepsilon_{0}$) in place of $\varepsilon_{\mathrm{eff}}$ in
the Wannier-Mott formula, respectively. In our study, the static dielectric
constant ($\varepsilon_{0}$) is computed by the formula, $\varepsilon_{0}=\varepsilon_{\infty}+\varepsilon_{ion}$.
From Table S2, it is evident that both $E_{Bu}$ and $E_{Bl}$ values
gradually decrease from P to Bi-containing systems, indicating stronger
dielectric screening with heavier pnictogens. Notably, the $E_{Bl}$
values are considerably lower than $E_{Bu}$ across all systems, reflecting
the dominance of electronic contribution over ionic contribution in
dielectric screeing. 

Moreover, a comparison of $E_{B}$ calculated using WM model and first-principles
method reveals that GW-BSE values (0.258$-$0.318\,eV) are consistently
higher than those predicted by the WM model (0.119$-$0.177 eV for
$E_{Bu}$ and 3$-$29 meV for $E_{Bl}$) as expected. Nevertheless,
both approaches show a consistent trend of decreasing $E_{B}$ from
P to Bi, indicating stronger dielectric screening in heavier pnictides.

Next, using the effective dielectric constant ($\mathrm{\varepsilon_{eff}}$)
and the reduced mass ($\mu^{\ast}$) of charge carriers, key excitonic
parameters such as exciton radius ($r_{exc}$), exciton lifetime ($\tau_{exc}$),
and the probability of a wavefunction ($|\phi_{n}(0)|^{2}$) for electron-hole
pair at zero separation is estimated (for detailes, see Supplemental
Material). As seen in Table \ref{tab:3}, the exciton radius ($r_{exc}$)
gradually increases from Ca$_{3}$PI$_{3}$ (0.88 nm) to Sr$_{3}$BiI$_{3}$
(1.15 nm), indicating more delocalized excitons in heavier pnictide
systems. This increasing delocalization corresponds to decreasing
$|\phi_{n}(0)|^{2}$ values implying a longer exciton lifetime in
Bi-containing compounds. Such extended lifetimes may benefit optoelectronic
performance by reducing carrier recombination. Overall, Sr-based systems
show slightly larger excitonic radii and lower $|\phi_{n}(0)|^{2}$
than their Ca counterparts, reflecting weaker Coulomb interaction
and enhanced dielectric screening. The exciton lifetime ($\tau_{exc}$)
is inversely proportional to $|\phi_{n}(0)|^{2}$ (for detailes, see
Supplemental Material). From Table \ref{tab:3} , it is evident that
$|\phi_{n}(0)|^{2}$ decreases from 0.05$\times$10$^{28}$ m$^{-3}$
in Ca$_{3}$PI$_{3}$ to 0.02$\times$10$^{28}$ m$^{-3}$ in Sr$_{3}$BiI$_{3}$
indicating $\tau_{exc}$ increases across the series, with heavier
B-site elements (Sb, Bi) and Sr-based compounds exhibiting longer
exciton lifetimes compared to their Ca counterparts. Such prolonged
lifetimes are beneficial for suppressing recombination losses and
improving the efficiency of optoelectronic devices.

Furthermore, the impact of phonon screening on exciton binding energy
($E_{B}$) is determined using the correction model recently proposed
by Filip et al. \citep{C2-5}. We determined the characteristic phonon
angular frequency ($\omega_{LO}$) using the thermal \textquotedbl B\textquotedbl{}
scheme proposed by Hellwarth et al. \citep{C1-30}, which involves
considering the spectral average of infrared-active optical phonon
modes (for details, see Supplemental Material). The correction due
to phonon screening is evaluated as:
\begin{center}
\begin{equation}
\ensuremath{\ensuremath{\Delta}}E_{B}^{ph}=-2\omega_{LO}\left(1-\frac{\varepsilon_{\infty}}{\varepsilon_{0}}\right)\frac{\sqrt{1+\omega_{LO}/E_{B}}+3}{\left(1+\sqrt{1+\omega_{LO}/E_{B}}\right)^{3}}
\end{equation}
\par\end{center}

\begin{flushleft}
As summarized in Table \ref{tab:3}, phonon screening leads marginal
reduction in $E_{B}$, ranging from 1.97 \% to 3.36 \%, with the largest
correction in Ca$_{3}$PI$_{3}$ and the smallest in Sr$_{3}$BiI$_{3}$.
This trend reflects slightly stronger screening effects in lighter
pnictide systems, whereas heavier Bi-based compounds show weaker corrections.
Overall, the reduction remains minor, indicating that electronic screening
is the dominant factor influencing excitonic behavior in X$_{3}$BI$_{3}$
systems.
\par\end{flushleft}

\begin{table}[H]
\caption{\label{tab:3}Calculated excitonic parameters for X$_{3}$BI$_{3}$
(X = Ca, Sr; B = P, As, Sb, Bi) antiperovskite derivatives. $E_{B}$
is the exciton binding energy calculated through BSE method, $r_{exc}$
is the exciton radius, $|\phi_{n}(0)|^{2}$ is the probability of
a wavefunction for electron-hole pair at zero separation, $\Delta E_{B}^{ph}$
is the phonon screening corrections of exciton binding energy, and
($E+\Delta E_{B}^{ph}$) is the corrected values of exciton binding
energy, respectively.}

\centering{}%
\begin{tabular}{cccccccc}
\hline 
Configurations &  & $E_{B}$ (eV) & $r_{exc}$ (nm) & $|\phi_{n}(0)|^{2}$ ($10^{28}$ $\mathrm{m^{-3}}$) & $\Delta E_{B}^{ph}$ (meV) & Reduction of $E_{B}$ (\%) & ($E_{B}+\Delta E_{B}^{ph}$) (meV)\tabularnewline
\hline 
Ca$_{3}$PI$_{3}$ &  & 0.305 & 0.88 & 0.05 & -10.25 & 3.36 & 294.75\tabularnewline
Sr$_{3}$PI$_{3}$ &  & 0.313 & 0.93 & 0.04 & -9.01 & 2.88 & 303.99\tabularnewline
Ca$_{3}$AsI$_{3}$ &  & 0.292 & 0.98 & 0.03 & -9.57 & 3.28 & 282.43\tabularnewline
Sr$_{3}$AsI$_{3}$ &  & 0.318 & 1.01 & 0.03 & -7.05 & 2.22 & 310.95\tabularnewline
Ca$_{3}$SbI$_{3}$ &  & 0.308 & 1.02 & 0.03 & -8.73 & 2.83 & 305.17\tabularnewline
Sr$_{3}$SbI$_{3}$ &  & 0.306 & 1.02 & 0.03 & -6.40 & 2.09 & 303.91\tabularnewline
Ca$_{3}$BiI$_{3}$ &  & 0.258 & 1.14 & 0.02 & -8.07 & 3.13 & 254.87\tabularnewline
Sr$_{3}$BiI$_{3}$ &  & 0.285 & 1.15 & 0.02 & -5.62 & 1.97 & 279.38\tabularnewline
\hline 
\end{tabular}
\end{table}

\subsection{\textit{Polaronic Properties:}}

Polaron formation originates from the interaction of charge carriers
with longitudinal optical (LO) phonons, which modifies their effective
transport behavior and plays a crucial role in determining the optoelectronic
potential of a material. This phenomenon is quantitatively described
by the Fr\"ohlich interaction model \citep{C1-29,C2-6}, which introduces
a dimensionless coupling parameter $\alpha$ as follows:
\begin{center}
\begin{equation}
\alpha=\left(\frac{1}{\varepsilon_{\infty}}-\frac{1}{\varepsilon_{0}}\right)\sqrt{\frac{R_{\infty}}{ch\omega_{LO}}}\sqrt{\frac{m^{*}}{m_{e}}}
\end{equation}
\par\end{center}

where $h$ is the Planck's constant, $c$ is the speed of light, and
$R_{\infty}$ denotes the Rydberg's constant. Typically, $\alpha\ll1$
reflects weak electron (or hole)-phonon coupling, while $\alpha>10$
correspond to strong coupling \citep{C1-13}. The computed $\alpha$
values (Table \ref{tab:4}) for the X$_{3}$BI$_{3}$ systems span
a moderate range from 1.91 to 4.83, indicating intermediate electron-phonon
coupling. This regime promotes delocalized large polarons, which are
beneficial for maintaining charge mobility without inducing excessive
localization.

To estimate the increase in carrier mass due to polaron formation,
we use Feynman\textquoteright s extended model \citep{C1-11,C2-32}:
\begin{center}
\begin{equation}
m_{p}=m^{*}\left(1+\frac{\alpha}{6}+\frac{\alpha^{2}}{40}+...\right)
\end{equation}
\par\end{center}

The calculated mass enhancements range from 1.47 to 2.39 for electrons
and 1.41 to 2.11 for holes with the largest increase observed in Ca$_{3}$PI$_{3}$
due to its stronger coupling. A general trend is observed where polaron
mass decreases from lighter (P) to heavier (Bi) pnictogens, which
aligns with the weakening of $\alpha$.

It is important to note that polaron formation can lead to a reduction
in the energies of electron and hole quasiparticles (QPs). This energy
lowering, referred to as the polaron energy ($E_{p}$), can be evaluated
from the coupling constant $\alpha$ using the following relation
\citep{C2-31,C2-34}:
\begin{center}
\begin{equation}
E_{p}=(-\alpha-0.0123\alpha^{2})\hbar\omega_{LO}
\end{equation}
\par\end{center}

\begin{flushleft}
For Ca$_{3}$PI$_{3}$, Sr$_{3}$PI$_{3}$, Ca$_{3}$AsI$_{3}$, Sr$_{3}$AsI$_{3}$,
Ca$_{3}$SbI$_{3}$, Sr$_{3}$SbI$_{3}$, Ca$_{3}$BiI$_{3}$, and
Sr$_{3}$BiI$_{3}$, the QP gap is reduced by 117.75, 98.49, 89.61,
75.26, 77.59, 65.27, 62.44, and 52.69 meV, respectively. Comparing
these values with $E_{B}$ from Table \ref{tab:3}, we conclude that
the charge-separated polaronic states are less stable than the bound
excitons \citep{C1-18}. Next, we calculated polaron mobilities ($\mu_{p}$)
for electrons and holes using the Hellwarth variational model \citep{C1-30}:
\begin{equation}
\mu_{p}=\frac{\left(3\sqrt{\pi}e\right)}{2\pi c\omega_{LO}m^{*}\alpha}\frac{\sinh(\beta/2)}{\beta^{5/2}}\frac{w^{3}}{v^{3}}\frac{1}{K(a,b)}
\end{equation}
\par\end{flushleft}

with
\begin{center}
\begin{equation}
K(a,b)=\int_{0}^{\infty}du\left[u^{2}+a^{2}-b\cos(vu)\right]^{-3/2}\cos(u)
\end{equation}
\par\end{center}

Here, $a^{2}$ and $b$ are determined as:
\begin{center}
\begin{equation}
a^{2}=(\beta/2)^{2}+\frac{(v^{2}-w^{2})}{w^{2}v}\beta\coth(\beta v/2)
\end{equation}
\begin{equation}
b=\frac{v^{2}-w^{2}}{w^{2}v}\frac{\beta}{\sinh(\beta v/2)}
\end{equation}
\par\end{center}

where $\beta=hc\omega_{LO}/k_{B}T$, $w$ and $v$ are temperature-dependent
parameters (for details, see Supplemental Material). Polaron mobility
values depicted in Table \ref{tab:4} lie between 7.23$-$37.19 cm$^{2}$V$^{-1}$s$^{-1}$,
with Sr$_{3}$BiI$_{3}$ exhibiting the highest $\mu_{p}$ for electrons
(26.23 cm$^{2}$V$^{-1}$s$^{-1}$) and Ca$_{3}$BiI$_{3}$ exhibiting
highest $\mu_{p}$ for holes (37.19 cm$^{2}$V$^{-1}$s$^{-1}$).
We also observe an increasing trend in $\mu_{p}$ from P to Bi, attributed
to reduced $\alpha$ and $m_{p}$. Compared to MAPbI$_{3}$, an well-known
halide perovskite, our systems exhibit lower mobilities \citep{C2-7},
yet remain competitive for optoelectronic use, particularly where
thermal stability and non-toxicity are prioritized. Overall, Bi-based
systems show improved mobility due to lower polaron mass and weaker
coupling, underscoring their promise for efficient charge transport
in lead-free photovoltaic applications.

\begin{table}[H]
\caption{\label{tab:4}Polaron parameters corresponding to electrons ($e$)
and holes ($h$) for X$_{3}$BI$_{3}$ (X = Ca, Sr; B = P, As, Sb,
Bi) antiperovskite derivatives.}

\centering{}%
\begin{tabular}{cccccccccccccc}
\toprule 
\multirow{2}{*}{Configurations} & \multirow{2}{*}{$\omega_{LO}$ (THz)} & \multirow{2}{*}{} & \multicolumn{2}{c}{$\alpha$} &  & \multicolumn{2}{c}{$m_{p}/m^{*}$} &  & \multicolumn{2}{c}{$E_{p}$ (meV)} &  & \multicolumn{2}{c}{$\mu_{p}$ (cm$^{2}$V$^{-1}$s$^{-1}$)}\tabularnewline
\cmidrule{4-5} \cmidrule{5-5} \cmidrule{7-8} \cmidrule{8-8} \cmidrule{10-11} \cmidrule{11-11} \cmidrule{13-14} \cmidrule{14-14} 
 &  &  & $e$ & $h$ &  & $e$ & $h$ &  & $e$ & $h$ &  & \multirow{1}{*}{$e$ } & $h$\tabularnewline
\midrule
Ca$_{3}$PI$_{3}$ & 3.00 &  & 4.45 & 4.83 &  & 2.24 & 2.39 &  & 56.37 & 61.38 &  & 9.84 & 7.23\tabularnewline
Sr$_{3}$PI$_{3}$ & 2.79 &  & 3.72 & 4.11 &  & 1.97 & 2.11 &  & 46.69 & 51.80 &  & 14.90 & 10.39\tabularnewline
Ca$_{3}$AsI$_{3}$ & 3.52 &  & 2.84 & 2.93 &  & 1.68 & 1.70 &  & 44.16 & 45.45 &  & 18.59 & 16.79\tabularnewline
Sr$_{3}$AsI$_{3}$ & 2.72 &  & 3.20 & 3.37 &  & 1.79 & 1.85 &  & 36.64 & 38.62 &  & 19.07 & 15.95\tabularnewline
Ca$_{3}$SbI$_{3}$ & 3.94 &  & 2.43 & 2.29 &  & 1.51 & 1.55 &  & 37.59 & 40 &  & 21.19 & 26.03\tabularnewline
Sr$_{3}$SbI$_{3}$ & 2.91 &  & 2.67 & 2.62 &  & 1.62 & 1.61 &  & 33.07 & 32.42 &  & 22.95 & 24.48\tabularnewline
Ca$_{3}$BiI$_{3}$ & 3.91 &  & 2.14 & 1.91 &  & 1.47 & 1.41 &  & 35.15 & 31.29 &  & 25.50 & 37.19\tabularnewline
Sr$_{3}$BiI$_{3}$ & 2.75 &  & 2.50 & 2.34 &  & 1.57 & 1.53 &  & 28.29 & 26.46 &  & 26.23 & 32.65\tabularnewline
\bottomrule
\end{tabular}
\end{table}

\section{Conclusions:}

In summary, our systematic investigation of the structural, electronic,
excitonic, and polaronic properties of X$_{3}$BI$_{3}$ (X = Ca,
Sr; B = P, As, Sb, Bi) systems using state-of-the-art first-principles
calculations unveils, all compounds are structurally and mechanically
stable. Dynamical stability was confirmed for five out of eight systems
at 0\,K through phonon calculations based on DFPT. These materials
exhibit direct G$_{0}$W$_{0}$@PBE bandgaps ranging from 2.42 to
3.02\,eV, ideal for efficient light-harvesting with minimal energy
loss. Exciton binding energy ranges between 0.258 and 0.318\,eV, ensuring
efficient exciton dissociation. A combined analysis using the BSE
and Wannier-Mott models confirms the dominant role of electronic screening,
with phonon screening contributing marginal corrections (1.97$-$3.36\%)
in $E_{B}$. Fr\"ohlich coupling constants ($\alpha$ = 2.14$-$4.83)
indicate weak-to-intermediate electron-phonon interaction, facilitating
large polaron formation. This is further supported by moderate polaron
masses and polaron energies. Notably, the polaron mobilities reach
up to 37.19 cm$^{2}$V$^{-1}$s$^{-1}$, highlighting the potential
of these materials for efficient charge transport. Taken together,
our findings suggest that X$_{3}$BI$_{3}$ compounds offer a promising
platform for lead-free, stable, and efficient optoelectronic applications.
\begin{acknowledgments}
A.C. would like to acknowledge the Shiv Nadar Institution of Eminence
(SNIoE) for funding and support. S.A. would like to acknowledge the
Council of Scientific and Industrial Research (CSIR), Government of
India {[}Grant No. 09/1128(11453)/2021-EMR-I{]} for Senior Research
Fellowship. The authors acknowledge the High Performance Computing
Cluster (HPCC) \textquoteleft Magus\textquoteright{} at SNIoE for
providing computational resources that have contributed to the research
results reported within this paper.
\end{acknowledgments}

\section*{DATA AVAILABILITY}

The data that support the findings of this article are not publicly
available. The data are available from the authors upon reasonable
request.

\bibliographystyle{apsrev4-2}
\bibliography{refs}

\end{document}